\documentstyle[sprocl]{article}

\input{psfig}
\def\Journal#1#2#3#4{{#1} {\bf #2}, #3 (#4)}

\def\NPB{{\em Nucl. Phys.} B}

\def\PRD{{\em Phys. Rev.} D}

\def\IJMPA{{\em Int. J. Mod. Phys.} A}

\def\be{\begin{equation}}
\def\ee{\end{equation}}
\def\bea{\begin{eqnarray}}
\def\eea{\end{eqnarray}}

\def\vev#1{\langle #1 \rangle}
\def\etal{{\it et al.}}

\def\mt{m_t}

\def\ntc{N_{TC}}

\def\gam{\gamma}
\def\lam{\lambda}

\def\br{B}
\def\lsim{\mathrel{\raise.3ex\hbox{$<$\kern-.75em\lower1ex\hbox{$\sim$}}}}
\def\gsim{\mathrel{\raise.3ex\hbox{$>$\kern-.75em\lower1ex\hbox{$\sim$}}}}

\def\mupmum{\mu^+\mu^-}
\def\epem{e^+e^-}
\def\tauptaum{\tau^+\tau^-}

\def\rts{\sqrt s}
\def\ie{{\it i.e.}}

\def\anti{\overline}

\def\mw{m_W}
\def\mz{m_Z}
\def\fbi{~{\rm fb}^{-1}}
\def\fb{~{\rm fb}}

\def\mev{~{\rm MeV}}
\def\gev{~{\rm GeV}}

\def\pzero{P^0}
\def\mpzero{m_{\pzero}}

\newcommand{\nn}{\nonumber}
\def\to{\rightarrow}

\begin{document}

\font\fortssbx=cmssbx10 scaled \magstep2
\hbox to \hsize{
%
%
$\vcenter{
\hbox{\fortssbx University of Florence}
}$
\hfill
$\vcenter{
\hbox{\bf DFF-346/9/99}
}$
}
\vskip1truecm

\title{THE PHENOMENOLOGY OF THE LIGHTEST PSEUDO NAMBU
GOLDSTONE BOSON AT FUTURE COLLIDERS
\footnote{Talk given  at the International 
Workshop on Linear Colliders, Sitges, Barcelona,
Spain, April 28-May 5, 1999.}}

\author{ D. DOMINICI}

\address{Dipartimento di Fisica, Universit\`a di Firenze and INFN,\\
Largo E. Fermi 2,
I-50125 Firenze, Italia}


\maketitle\abstracts{
The capability of the linear collider to discover and study
the lightest neutral pseudo-Nambu-Goldstone
boson ($\pzero$) of dynamical symmetry breaking models
in the $e^+e^-$ and $\gam\gam$ modes is  presented. 
For a number of technicolor $\ntc=4$, the
discovery of the $\pzero$ at an $\epem$ collider
via the reaction $\epem\to\gam \pzero$ should
be possible for an integrated luminosity of $L=100\fbi$ at $\rts=500\gev$
as long as $\mpzero$ is not near $\mz$. 
In the $\gam\gam$ collider mode
 the $\gam\gam\to\pzero\to b\anti b$ signal should
be very robust and could be measured with high statistical
accuracy for a broad range of $\mpzero$ if $\ntc=4$.}

\section{Introduction}
Theories of the electroweak interactions based on dynamical
symmetry breaking (DSB) avoid the introduction of
fundamental scalar fields but generally predict
many pseudo-Nambu-Goldstone bosons (PNGB's)
due to the breaking of a large initial global symmetry group $G$.
Among the PNGB's the colorless neutral states are the lightest ones.
Direct observation of a PNGB would not have
been possible at any existing accelerator,
however light the PNGB's are, unless
the number of technicolors, denoted $\ntc$, is very large.
The phenomenological analysis presented here is extracted  
from ref.\cite{pgb},
where all the details can be found, 
and is based    on a  $SU(8)\times SU(8)$ 
effective low-energy Lagrangian approach.
 In the broad class
of models considered, the lightest neutral PNGB $\pzero$ is of 
particular interest
because it contains only down-type techniquarks (and charged technileptons)
and thus will have a mass scale that is most naturally  set by the
mass of the $b$-quark. 
The $\pzero$ total width is typically in the few $\mev$  range
and dominant decay modes are 
$b\anti b$, $\tau^+\tau^-$ and $gg$.
Other color-singlet PNGB's will have masses
most naturally set by $\mt$, while color non-singlet PNGB's will
generally be even heavier. 

Detection of the PNGB's at the Tevatron and LHC colliders,
has been extensively considered~\cite{chivu}.
However, inclusive $gg$ fusion production of a neutral PNGB,
followed by its decay to $\gam\gam$, was not given detailed consideration
until recently~\cite{pgb}.  In this paper it was noticed that for
a particular class of models the ratio $\Gamma(\pzero\to gg)B(\pzero\to
\gam\gam)/\Gamma(H\to gg)B(H\to\gam\gam)$ with $H$ being the SM Higgs
and $N_{TC}=4$ 
is of the order $10^2$ for $50 \leq m_{\pzero/H}(\gev)\leq 150$.
Therefore,  using the results on the Higgs analysis, we can   conclude
that, for $\ntc = 4$,
the $\pzero$ can be detected in the $gg\to\pzero\to\gam\gam$ mode
for at least $30 - 50<\mpzero<150 - 200\gev$,
or perhaps also at Tevatron RunII with $S/\sqrt{B}\geq 3$ for
$m_{\pzero}\geq 60\gev$.

\section{$\epem$ mode }

The best mode for $\pzero$ production
at an $\epem$ collider (with $\rts>\mz$) is $\epem\to\gam\pzero$.
Because the $\pzero Z\gam$ coupling-squared is much smaller than the
$\pzero\gam\gam$ coupling-squared,
the dominant diagram is $\epem\to\gam\to\gam\pzero$.
 Even when kinematically allowed,
rates in the $\epem\to Z\pzero$ channel are substantially smaller,
as we shall discuss. We will give results for the moderate
value of $\ntc=4$.
For $\rts=200\gev$, we find that,
after imposing an
angular cut of $20^\circ\leq\theta\leq 160^\circ$ on the outgoing
photon (a convenient acceptance cut
that also avoids the forward/backward cross section
singularities but is more than 91\% efficient),
the $\epem\to\gam\pzero$ cross section is below $1\fb$ for $\ntc=4$.
Given that the maximum integrated luminosity anticipated
is of order $L\sim 0.5\fbi$, we conclude that LEP2
will not allow detection of the $\pzero$ unless $\ntc$ is very large.

The cross section for $\epem\to \gam \pzero$ at $\rts=500\gev$,
after imposing the same angular  cut,  ranges from
$0.9\fb$ down to $0.5\fb$ as $\mpzero$ goes from zero up to $\sim 200\gev$.
For $L=50\fbi$, we have at most 45 events with which to discover
and study the $\pzero$.
The $\epem\to Z\pzero$ cross section is even smaller. Without cuts
and without considering any specific $Z$ or $\pzero$ decay modes, it
ranges from $0.014\fb$ down to $0.008\fb$ over the same mass range.
  If TESLA is able to achieve
$L=500\fbi$ per year, $\gam\pzero$ production will have a substantial
rate, but the $Z\pzero$ production rate will still not be useful.
Since the $\gam\pzero$ production rate scales as $\ntc^2$, if $\ntc=1$
a $\rts=500\gev$ machine will yield at most 3 (30) events for
$L=50\fbi$ ($500\fbi$), making $\pzero$ detection and study extremely
difficult.  Thus, we will focus our analysis on the $\ntc=4$ case.

In order to assess the $\gam\pzero$ situation more fully,
we must consider backgrounds.
The dominant decay modes of the $\pzero$ are  typically to
$b\anti b$, $\tauptaum$ or $gg$.  For the $b\anti b$ and $gg$
modes, the backgrounds relevant to the $\gam \pzero$
channel are $\gam b\anti b$, $\gam c\anti c$ and $\gam q\anti q$
($q=u,d,s$) production. The cross sections for these processes
obtained after integrating over a 10 GeV bin size in the quark-antiquark
mass  are, for $10\lsim\mpzero\lsim 80\gev$
and $\mpzero\geq 100\gev$, of the same order of the  signal.

\begin{figure}[htb]
\centerline{
\psfig{figure=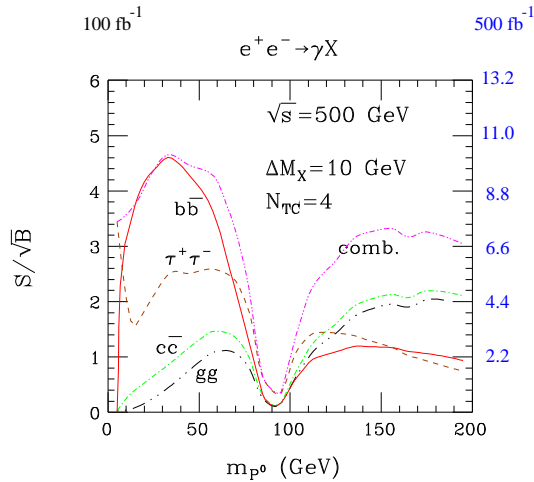,height=2.5in}}
\caption{The statistical significances 
$S/\protect\sqrt B$ for a $\pzero$ signal
in various `tagged' channels as a function
of $\mpzero$ at a $500\gev$ collider for  integrated luminosities 
of $100\fbi$ and $500\fbi$.
\label{fig:eensd}}
\end{figure}

Results for $S/\sqrt B$, in the various tagged channels, for $\ntc=4$ and
assuming $L=100\fbi$ (and $L=500\fbi$) at $\rts=500\gev$,
are plotted in Fig.~\ref{fig:eensd}.
We have assumed a mass window of $\Delta M_X=10\gev$ in evaluating
the backgrounds in the various channels. Also shown
in Fig.~\ref{fig:eensd}
is the largest $S/\sqrt B$ that can be achieved by considering
(at each $\mpzero$) all possible combinations of the $gg$, $c\anti c$,
$b\anti b$ and $\tauptaum$ channels.  From the figure, we find
for $L=100\fbi$
$S/\sqrt B\geq 3$ (our discovery criterion)
for $\mpzero\leq 75\gev$ and $\mpzero\geq 130\gev$, \ie\
outside the $Z$ region.  A strong signal, $S/\sqrt B\sim 4$, is only
possible for $\mpzero\sim 20-60\gev$. As the figure shows, the signal
in any one channel is often too weak for discovery, and it is only
the best channel combination that will reveal a signal.
For the TESLA $L=500\fbi$ luminosity, $S/\sqrt B$ should be multiplied
by $\sim 2.2$ and discovery prospects will be improved.
Tagging and mistagging efficiencies have been included \cite{pgb}.
 
After discovery, one can determine branching fractions in various channels and
couplings. The only channel with reasonable ($\leq 15\%$) statistical
error would be $b\anti b$, for $L=500\fbi$.

\section{$\gam\gam$ mode }

By folding the cross section 
for the $\pzero$ production at a given energy $E_{\gam\gam}$
of  a $\gam\gam$ collider with the  differential luminosity,
one gets  \cite{gunhabgamgam}
\bea
N(\gam\gam\to \pzero \to F)&=& {8\pi\Gamma(\pzero\to\gam\gam)
\br(\pzero\to F)
\over m_{\pzero}^2E_{\epem}}\tan^{-1}{\Gamma_{\rm exp}\over 
\Gamma^{\rm tot}_{\pzero}}\nn\\
&&\times
\left(1+\vev{\lam\lam^\prime}\right)G(y_0)L_{\epem}\,,
\label{siggamgam}
\eea
where $y_0= m_{\pzero}/E_{\epem}$, $\lam$ and $\lam^\prime$ 
are the helicities of the colliding photons,
$\Gamma_{\rm exp}$ is the mass interval accepted in the final state
$F$ and $L_{\epem}$ is the integrated luminosity for the colliding
electron and positron beams. For initial discovery one chooses initial
laser polarizations $P$ and $P^\prime$ and $\epem$   beam helicities
$\lam_e$ and $\lam_e^\prime$  for a broad spectrum 
$2\lam_e P\sim +1$, $2\lam_e^\prime P^\prime\sim+1$,
$PP^\prime\sim +1$ such that $G\gsim 1$ and $\vev{\lam\lam^\prime}\sim 1$
(which suppresses $\gam\gam\to q\anti q$ backgrounds)
over the large range $0.1\leq y_0\leq 0.7$. 
The $\pzero$ is always
sufficiently narrow that $\tan^{-1}\to \pi/2$. In this limit,
the rate is proportional to $\Gamma(\pzero \to\gam\gam)\br(\pzero\to F)$.
For the $\pzero$, $\Gamma(\pzero\to\gam\gam)$ is large and the total
production rate will be substantial. 

\begin{figure}[htb]
\centerline{
\psfig{figure=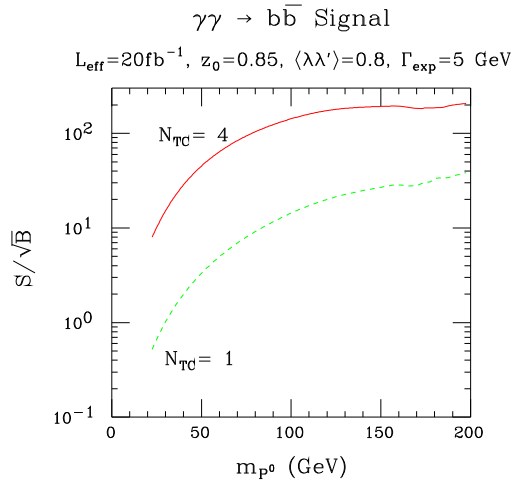,height=2.5in}}
\caption{The statistical significance $S/\protect\sqrt B$ for $\ntc=4$
and $\ntc=1$ at a $500\gev$ $\gam\gam$ collider.
\label{fig:gamgamnsd}}
\end{figure}

 Since it is
well-established \cite{gunhabgamgam,baueretal,japanese} that the SM $h$
can be discovered in this decay mode for $40\lsim m_h\lsim 2\mw$, 
it is clear
that $\pzero$ discovery in the $b\anti b$
final state will be possible up to at least $200\gev$, down to
$\sim 0.1 \rts\sim 50\gev$ (at $\rts\sim 500\gev$), below which $G(y)$
starts to get small. Discovery at lower values of $\mpzero$ would
require lowering the $\rts$ of the machine.
For the $b\anti b$ channel, the statistical significance 
 $S/\sqrt B$ is  plotted in Fig.~\ref{fig:gamgamnsd}.

Once the $\pzero$ has been discovered, either in $\gam\gam$ collisions
or elsewhere, one can configure the $\gam\gam$ collision set-up so
that the luminosity is peaked at $\rts_{\gam\gam}\sim \mpzero$.
A very precise measurement of the $\pzero$ rate in the $b\anti b$
final state will then be possible if $\ntc=4$.
For example, rescaling the SM Higgs `single-tag'
results of Table 1 of Ref.~\cite{japanese} (which assumes
a peaked luminosity distribution with a total of $L=10\fbi$)
for the $106\gev\leq m_{jj}\leq 126\gev$ mass window to the case of
the $\pzero$, 
we obtain $S\sim 5640$ compared to $B\sim 325$,
after angular, topological tagging and jet cuts. This implies a statistical
error for measuring $\Gamma(\pzero\to \gam\gam)\br(\pzero\to b\anti b)$
of $\lsim 1.5\%$. Systematic errors will probably dominate.
Following the same procedure for $\ntc=1$, we find (at this mass) a
statistical error for this measurement of $\lsim 5\%$. Of course,
for lower masses the error will worsen.
For $\ntc=4$, we estimate an error for the $b\anti b$ rate
measurement still below $10\%$ even at a mass as low
as $\mpzero=20\gev$ (assuming the $\rts$ of the machine
is lowered sufficiently to focus on this mass without sacrificing luminosity).
For $\ntc=1$, we estimate an error for the $b\anti b$ rate measurement
of order $15-20\%$ for $\mpzero\sim 60\gev$.

\section{Conclusions}
We have reviewed  the production and study of the
lightest pseudo-Nambu Goldstone state  $\pzero$
of a typical technicolor model at future  colliders, focusing
mainly on $\epem$. 
For $\ntc=4$, discovery of the $\pzero$ in the $gg\to\pzero\to\gam\gam$
mode at the LHC will be almost certainly be possible unless its
mass is either very small ($\lsim 30\gev$?) or very large ($\gsim
200\gev$?), where the question marks are related to uncertainties in LHC
backgrounds in the inclusive $\gam\gam$ channel.

In contrast, an $\epem$ collider, while able to discover the $\pzero$
via $\epem\to\gam\pzero$, so long as $\mpzero$ is not close to $\mz$
and $\ntc\geq 3$,
is unlikely (unless the TESLA $500\fbi$ per year option
is built or $\ntc$ is very large) to be able to determine the rates for
individual $\gam F$ final states ($F=b\anti b,\tauptaum,gg$
being the dominant $\pzero$ decay modes) with sufficient accuracy as
to yield more than very rough indications regarding the important
parameters of the technicolor model.

The $\gam\gam$  option at an $\epem$ collider  is actually a more
robust means for discovering the $\pzero$ than direct operation
in the $\epem$ collision mode.  For $\ntc=4$,
$\gam\gam\to\pzero\to b\anti b$
should yield an easily detectable $\pzero$ signal for
$0.1\lsim{\mpzero\over\rts}\lsim 0.7$.  Once
$\mpzero$ is known, the $\gam\gam$ collision set-up can be re-configured
to yield a luminosity distribution that is strongly peaked at
$\rts_{\gam\gam}\sim \mpzero$ and,
for much of the mass range of $\mpzero\lsim 200\gev$,
a measurement of $\Gamma(\pzero\to \gam\gam)\br(\pzero\to b\anti b)$
can be made with statistical accuracy in the $\lsim 2\%$ range.

A $\mupmum$ collider would be crucial for detecting a light
$\pzero$ ($\mpzero\lsim 30\gev$)
 and would play a very special role
with regard to determining key properties of the $\pzero$ 
\cite{pgb}.
In particular, the $\pzero$, being, in the class 
of models we have considered,
 comprised
of $D\anti D$ and $E\anti E$ techniquarks, will naturally have couplings
to the down-type quarks and charged leptons of the SM.
Thus, $s$-channel production ($\mupmum\to\pzero$) is predicted
to have a substantial rate for $\rts\sim\mpzero$.
Because the $\pzero$ has a very narrow width, in order to maximize this rate
it is important that one operates the $\mupmum$ collider
so as to have extremely small beam energy spread, $R=0.003\%$.
The complete analysis of how 
the precision $\mupmum$ measurements of various channel
rates together with LHC and $\epem$ measurement
can determine (up to a discrete set of ambiguities)
the  parameters of the effective
low-energy Yukawa Lagrangian that determine $T_3=-1/2$
fermion masses and their couplings to the $\pzero$ can be 
found in \cite{pgb}.
 
\section*{Acknowledgments}
I would like to thank 
R. Casalbuoni, S. De Curtis, A. Deandrea, R. Gatto and J. Gunion
for the fruitful and enjoyable collaboration on the topics covered here
and R. R\"uckl for interesting discussions..

\end{document}